\documentclass[aps,prl,twocolumn,showpacs,amssymb,amsmath]{revtex4}

\usepackage{graphicx}

\begin{document}

\title{Efficient optical quantum state engineering}

\author{Kevin T. McCusker}
\email{kmccusk3@illinois.edu}
\author{Paul G. Kwiat}
\affiliation{University of Illinois at Urbana-Champaign}

\date{\today}

\begin{abstract}
We discuss a novel method of efficiently producing multi-photon states using repeated spontaneous parametric downconversion. Specifically, by attempting downconversion several times, we can pseudo-deterministically add photons to a mode, producing various several-photon states. We discuss both expected performance and experimental limitations.
\end{abstract}


\maketitle

Despite many advances in optical quantum information processing, such as quantum cryptography \cite{cryptography}, teleportation \cite{teleportation}, and quantum computing (both proposals \cite{qcomptheory} and simple demonstrations \cite{qcompexperiment}), reliably and deterministically creating even simple quantum optical states remains a challenge. For example, on-demand single-photon production is still elusive, despite significant recent progress \cite{singlephotonsource}. More complicated states can often be created probabilistically using single photons, linear optics, and feed-forward, but these schemes typically scale exponentially poorly with the number of photons in the state \cite{currentmethods} or are prohibitively complicated \cite{qcomptheory}. In this letter, we propose a novel method using repeated spontaneous parametric downconversion to closely approximate applying the creation operator, allowing efficient pseudo-deterministic preparation of a variety of states, with critical implications for applications including quantum computing and quantum metrology.

Spontaneous parametric downconversion (a nonlinear optical process in which one high energy photon in a laser beam splits into a pair of lower energy photons, called the signal and idler) has for many years been the workhorse for producing high quality simple photon states, such as heralded single photons \cite{heraldedphotons} and entangled photons. More recently, four-wave mixing (FWM) has been used to produce these states as well \cite{FWM,FWM2}. Downconversion has also been used to add a photon to a classical light field, with nonclassical results \cite{addsubphoton}. However, one of the drawbacks of these approaches is that they are nondeterministic, i.e., the number of pairs of photons that are produced is described by a random (thermal) distribution. One way to overcome this problem and produce single photons deterministically is to monitor the signal mode of several downconversion sources \cite{deterministicsps2}, or a single source pulsed at several times \cite{deterministicsps1,deterministicsps3}; this allows one to herald the output of the idler mode without directly measuring it, and then select the source/pulse that produced the desired output. We propose modifying this technique to drive downconversion (or FWM) weakly in a cavity until we produce exactly one pair, thereby deterministically adding a photon to the idler mode. Repeating this process, we can ``build up'' a desired number state. Furthermore, by manipulating the polarization of the photon that is being added (or equivalently, manipulating the polarization of the photons already created before the next one is added), we can efficiently and with high fidelity produce any state that is expressible as a product of creation operators of arbitrary polarization on a single mode:
\begin{eqnarray}
\label{productformula}
|\psi\rangle &= \displaystyle\prod_{n=0}^{N-1}(\alpha_n{a}^\dagger_H+\beta_n{a}^\dagger_V)|0\rangle.
\end{eqnarray}

\begin{figure}
\centering
\includegraphics[width=3.4in]{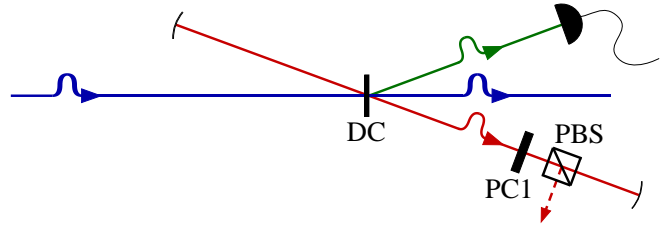}
\caption{Diagram of proposed Fock-state source. A pulsed laser pumps a downconversion crystal (DC). The signal photon of each created pair is detected, and the idler photon is emitted into a storage cavity. Photons are allowed to accumulate in the cavity until the desired number is reached. The light can be switched out by rotating its polarization with a Pockels cell (PC1), so the polarizing beam splitter (PBS) reflects rather than transmits it.}\label{diagram}
\end{figure}

First we will discuss the simplest application, creating Fock (photon-number) states. Our proposed design (somewhat similar to that discussed in \cite{condeng}) is shown in Fig.~\ref{diagram}. A series of laser pulses is incident on a nonlinear crystal, with each pulse resulting in some probability of producing one (or more) pair(s) of photons into two separate spatial modes (the signal and idler modes). The signal mode is detected, while the idler is allowed to propagate through a cavity. The cavity length is such that when the idler light makes one complete pass through the cavity and returns to the crystal, the next laser pulse is also passing through the crystal. This allows us to keep adding identical photons nondeterministically to the idler mode until we have the desired number of photons (as indicated by the total number of signal photons detected). Once we do, we release them using an optical switch, e.g., a Pockels cell and polarizing beam splitter (PBS). If we have a photon-number-resolving detector for the signal arm, we can add more than one photon on each pass, allowing us to build up the state in fewer passes (although we need to be careful not to overshoot the desired number of photons in the cavity).

\begin{figure}
\centering
\includegraphics[width=3.4in]{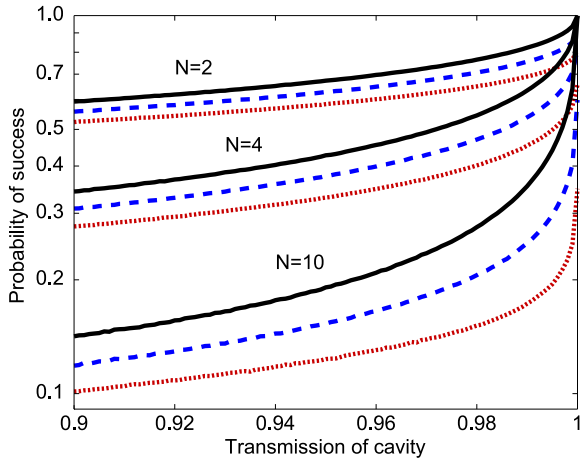}
\caption{Theoretical performance of Fock-state source versus cavity transmission. Curves are shown for several values of \textit{N} (labeled), and several detector efficiencies (black solid, $\eta$=1; blue dashed, $\eta$=0.95; red dotted, $\eta$=0.9).}\label{performance}
\end{figure}

Fig.~\ref{performance} shows the predicted performance as a function of cavity transmission, for different detector efficiencies. Here we assume a photon-number resolving detector \cite{detectors,VLPCs}, and the ability to tune the pump pulse intensity--and hence the expected number of pairs--for each pass at downconversion, although producing several pure pairs per pulse remains experimentally challenging \footnote{For states with large numbers of photons, effects similar to gain guiding may appear \cite{gainguiding}, where the parts of the idler beam that are relatively high intensity change the transverse probability distribution of the stimulated emission.} (without these assumptions, we would simply need more passes to prepare the desired number of photons). Additional experimental limitations, such as imperfect single-mode collection or undesirable frequency entanglement, are discussed below. Our predicted performance greatly exceeds current methods: e.g., creating a heralded four-photon Fock state via single-pass downconversion is in principle limited to 6.7$\%$ probability, and the best experimental result is only 0.2\% (with fidelity=0.6) \cite{dcfock}; even postselecting on an attenuated coherent source cannot do better than 19.6$\%$ probability. Our scheme could realistically produce this state with $>$50$\%$ probability.

\begin{figure}
\centering
\includegraphics[width=3.4in]{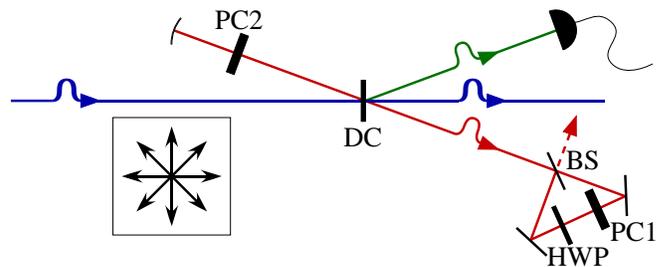}
\caption{Diagram of proposed \textit{N}00\textit{N}-state source. Similar to the setup for the Fock-state source (Fig.~\ref{diagram}), with the addition of a Pockels cell (PC2) in the cavity to rotate the polarization of the photons as they are created, and a polarization-independent switch (made up of a beamsplitter (BS), half-wave plate (HWP), and Pockels cell (PC1)) in place of a polarization-dependent switch. Inset shows the linear polarization of 4 photons in the desired state.}
\label{noondiagram}
\end{figure}

We now consider construction of more complicated states. One of the more interesting states that we can create is a number-path entangled state of the form $|N_{A},0_{B}\rangle+|0_{A},N_{B}\rangle$ (known as a ``\textit{N}00\textit{N}'' state \cite{precisemeasure}). A \textit{N}00\textit{N} state can be used to reach the Heisenberg limit for precision measurements, achieving a phase uncertainty that scales as 1/\textit{N} \cite{precisemeasure,precisemeasureexp1,precisemeasureexp2,precisemeasureexp3}. This same state can also be used for quantum lithography \cite{lithography}, demonstrating ``super resolution''. Originally proposed methods \cite{currentmethods,currentmethods2} for creating \textit{N}00\textit{N} states using linear optics scaled exponentially poorly with increasing \textit{N}, even assuming perfect optics, on-demand Fock-state sources, and detectors. A recent proposal \cite{recentproposal} suggests a method for creating \textit{N}00\textit{N} states that scales efficiently using linear optics and feed-forward, but the number of photons making up the \textit{N}00\textit{N} state varies nondeterministically in each attempt.

We start with the observation that a \textit{N}00\textit{N} state in the right/left circular polarization basis can be expressed as a product of linearly polarized photons (neglecting normalization) \cite{precisemeasureexp3}:
\begin{eqnarray}
\label{noonformula}
(\hat{a}^\dagger_R)^N-(\hat{a}^\dagger_L)^N =
\displaystyle\prod_{n=0}^{N-1}[cos(n\pi/N)\hat{a}^\dagger_H+sin(n\pi/N)\hat{a}^\dagger_V].
\end{eqnarray}
This state is the product of N photons superimposed on each other, with the polarization of the photons evenly spaced by 180$^\circ$/N (see Fig.~\ref{noondiagram} inset). We can construct this state by adding N photons one at a time to the field in the cavity, and rotating the polarization of all photons in the cavity by 180$^\circ$/N every time a new photon is added. The proposed setup, shown in Fig.~\ref{noondiagram}, is similar to the setup for making Fock states, with the addition of a Pockels cell to rotate the polarization of the light in the cavity \footnote{For the cavity design shown in Fig.~\ref{noondiagram}, it is experimentally simpler to create a \textit{N}00\textit{N} state in the 45$^\circ$/-45$^\circ$ basis, expressible as a product of elliptical polarization states: $(\hat{a}^\dagger_{45^\circ})^N-(\hat{a}^\dagger_{-45^\circ})^N = \prod_{n=0}^{N-1}[cos(n\pi/N)\hat{a}^\dagger_H+isin(n\pi/N)\hat{a}^\dagger_V].$ This is a topographically similar distribution on the Poincar\'{e} sphere as Eq.~(\ref{noonformula}), but in the vertical instead of horizontal plane. This distribution can be created using PC2 to incrementally rotate the initially horizontally polarized photons around 45$^\circ$ on the Poincar\'{e} sphere.}, and a polarization-independent switch \footnote{The switch can be realized with a Sagnac interferometer with a half-wave plate at 0$^\circ$ and a Pockels cell able to act as a half-wave plate at 45$^\circ$ (PC1). When PC1 is on, there is a 180$^\circ$-phase shift between the path that goes through the HWP then PC1 and the path that goes through PC1 then the HWP, allowing for on-demand switch-out.}. After the switch-out, wave plates and a polarizing beam splitter can convert the state to the desired number-path entangled state.

\begin{figure}
\centering
\includegraphics[width=3.4in]{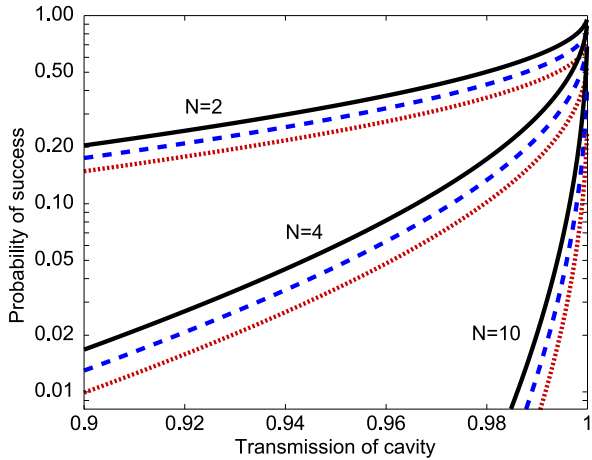}
\caption{Theoretical performance of \textit{N}00\textit{N}-state source. Curves are shown for several values of \textit{N} (labeled), and for several different detector efficiencies (black solid, $\eta$=1; blue dashed, $\eta$=0.95; red dotted, $\eta$=0.9).}
\label{performancenoon}
\end{figure}

The predicted performance is shown in Fig.~\ref{performancenoon}.
Comparing with Fig.~\ref{performance}, we can see that the probability of successfully producing a \textit{N}00\textit{N} state is significantly lower than that of producing a Fock state with the same number of photons. The primary reason is that the \textit{N}00\textit{N} state must be built up exactly one photon at a time, whereas for the Fock state several photons can be added in one pass. The additional passes for \textit{N}00\textit{N} state creation increase the sensitivity to cavity loss. Also, the fidelity of the produced state is not perfect due to higher-order terms in the downconversion Hamiltonian (see below). Nevertheless, our predicted performance exceeds current state-of-the-art experiments \cite{precisemeasureexp3} (by more than an order of magnitude) and previous proposals using linear optics. For example, the highest probability \cite{currentmethods2} of creating an \textit{N}=6 \textit{N}00\textit{N} state with linear optics is 0.097, assuming perfect on-demand Fock-state sources (12 photons total), perfect optics, and perfect detector efficiency. This level of performance is feasible with our proposal, \textit{without} these assumptions. Recent research shows that \textit{N}00\textit{N} states decohere very rapidly in the presence of loss \cite{fragile}; however, a similar superposition (of the form $|m,m'\rangle_{A,B}+|m',m\rangle_{A,B}, m>m'$) can greatly improve the robustness against decoherence while keeping the ability to perform sub-shot noise phase estimation \cite{mnm,mnm2}. To create such a state, we first observe that it can be expressed as the product of a \textit{N}00\textit{N}- and Fock-state creation operators:
\begin{eqnarray}
&&(\hat{a}^\dagger_A)^m(\hat{a}^\dagger_B)^{m'}+(\hat{a}^\dagger_A)^{m'}(\hat{a}^\dagger_B)^m=\nonumber\\
&&\Bigl( (\hat{a}^\dagger_A)^{m-m'}+(\hat{a}^\dagger_B)^{m-m'} \Bigr) (\hat{a}^\dagger_A)^{m'}(\hat{a}^\dagger_B)^{m'}.
\label{mnmformula}
\end{eqnarray}
Hence, in order to create this new state, we can create two Fock states in the cavity, and then add a \textit{N}00\textit{N} state.

We now briefly discuss the proposals limitations. For our performance plots in Figs.~\ref{performance} and~\ref{performancenoon}, a ``success'' is defined as an attempt with each created pair collected and detected, with no photons leaking out of the cavity before the process is complete, and with no extra photons (although for creating Fock states, it is still a success if the number of photons lost is equal to the number of extra photons). Photon loss for each pass through the cavity must therefore be minimized \footnote{If the cavity loss is polarization dependent, it can affect the state even if no photon is lost, as the loss acts as a partial polarizer \cite{precisemeasureexp3}.}, and the downconversion photons must be efficiently collected in pure states, i.e., they must be indistinguishable in every degree of freedom \cite{indistinguishable,frequnen}; similar requirements hold, e.g., to realize teleportation \cite{teleportation}.

The effect of the higher-order terms of the downconversion Hamiltonian must be taken into account. There will be some probability of creating two pairs, although this term can be completely eliminated if the transmission of the cavity is high enough, since in the limit of no cavity loss, the crystal could be pumped infinitely weakly, low enough that \textit{N} single pairs will almost certainly appear before one double pair is generated. Even when no pairs are created, the effect of higher-order terms in the Hamiltonian of the downconversion can alter the state in the cavity. Treating the pump pulse classically, we have \cite{Hamiltonian}:
\begin{eqnarray}
&e^{i\epsilon\hat{H}} = 1-\epsilon\hat{a}^{\dagger}\hat{b}^{\dagger}+\frac{\epsilon^2}{2}\hat{a}^{{\dagger}2}\hat{b}^{{\dagger}2}-\frac{\epsilon^2}{2}\hat{a}\hat{a}^{\dagger}\hat{b}\hat{b}^{\dagger},&
\label{exphamil}
\end{eqnarray}
where \textit{$\epsilon$} is the effective interaction strength, and \textit{$\hat{a}$} and \textit{$\hat{b}$} refer to the idler and signal modes, respectively. Terms of order \textit{$\epsilon^3$} are dropped, as are terms where \textit{$\hat{b}$} would be acting on the vacuum (giving zero). The second term of Eq.~(\ref{exphamil}) creates the desired single pair of photons. The third term creates an undesirable two pairs, which could be detected with a photon-number-resolving detector, and eliminated by driving weakly enough. The fourth term, which can be interpreted as the creation and then destruction of a pair, can alter the state in the cavity, even though it does not add or remove any photons. If, for example, we are trying to create a \textit{N}00\textit{N} state with \textit{N}=4, after the creation and rotation of two photons, the state in the cavity will be (neglecting normalization)
\begin{equation}
(\hat{a}^{\dagger}_H+\hat{a}^{\dagger}_V)\hat{a}^{\dagger}_V|0_H0_V\rangle = |1_H1_V\rangle+\sqrt{2}|0_H2_V\rangle.
\label{aftertwo}
\end{equation}
Applying the Hamiltonian in Eq.~(\ref{exphamil}) (assuming no signal photon is present, i.e., projecting out the contribution from the second and third terms), gives
\begin{eqnarray}
&&(1-\frac{\epsilon^2}{2}\hat{a}_H\hat{a}^{\dagger}_H)(|1_H1_V\rangle+\sqrt{2}|0_H2_V\rangle)=\nonumber\\
&&(1-\epsilon^2)|1_H1_V\rangle+(1-\frac{\epsilon^2}{2})\sqrt{2}|0_H2_V\rangle,
\label{afterpass}
\end{eqnarray}
which differs from the initial state in Eq.~(\ref{aftertwo}).  This change adds \textit{coherently} with each pass, and lowers the fidelity between the produced state and the desired state, even as $\epsilon$ approaches zero. However, the effective downconversion operator in Eq.~(\ref{afterpass}) can be undone (to order $\epsilon^2$) by applying the same effective operator with the orthogonal polarization:
\begin{eqnarray}
(1-\frac{\epsilon^2}{2}\hat{a}_H\hat{a}^{\dagger}_H)(1-\frac{\epsilon^2}{2}\hat{a}_V\hat{a}^{\dagger}_V)|\psi\rangle=\nonumber\\
(1-\frac{\epsilon^2}{2}(\hat{a}_H\hat{a}^{\dagger}_H+\hat{a}_V\hat{a}^{\dagger}_V))|\psi\rangle.
\label{afterdoublepass}
\end{eqnarray}
Since the state in the cavity $|\psi\rangle$ always has a definite number of photons, it is an eigenstate of $\hat{a}_H\hat{a}^{\dagger}_H+\hat{a}_V\hat{a}^{\dagger}_V$, and therefore an eigenstate of the operator in Eq.~(\ref{afterdoublepass}). We can approximate the alternate application of these operators by adding the photons to the cavity in a different order, resulting in a higher average overlap with the desired state (e.g., from about 0.56 to 0.83 for N=8). Preliminary results indicate that this state, with a photon distribution on the Poincar\'{e} sphere similar to that of a \textit{N}00\textit{N} state, is still useful for quantum metrology; more detailed investigation of this is an interesting possibility for future study.

In conclusion, we have proposed a novel technique that can efficiently produce a variety of multi-photon states with high fidelity, including Fock and \textit{N}00\textit{N} states. Although we discussed only the case where we start with the vacuum in the idler mode and build up states with a well-defined number of photons, these techniques can also be applied to states that do not have well-defined photon numbers, such as squeezed or coherent states. Another possibility which may allow for the creation of additional interesting states is supplying something other than the vacuum for the initial \textit{signal} field, such as a weak coherent state \cite{condeng} or zero-one photon entangled state \cite{scissors}. Finally, if we replace the weak downconversion source in our cavity with a weak beam splitter, and allow a state to pass through it multiple times until a photon is detected in the reflected path, we can in principle remove a single photon from the state with arbitrarily high efficiency. The ability to subtract photons allows for generation of interesting states, including \textit{N}00\textit{N} states \cite{noonbysub}. Combining the ability to both add and subtract photons \cite{addsubphoton} may allow for direct tests of fundamental physics (such as the bosonic commutation relation \cite{singlephotoninterference}) as well as creation of otherwise-unreachable states.

We thank A. Steinberg for his input. This work was supported by the MURI Center for Photonic Quantum Information Systems (IARPA program DAAD19-03-1-0199) and the Dept. of Interior (\#NBCHC070006).

\bibliographystyle{apsrev}

\end{document}